\documentclass[seceq]{ptptex}

\usepackage{wrapft} 
\usepackage[dvips]{graphicx}

\newcommand{\eqb}{\begin{equation}}
\newcommand{\eqe}{\end{equation}}
\newcommand{\eqbnon}{\begin{equation*}}
\newcommand{\eqenon}{\end{equation*}}

\newcommand{\eqab}{\begin{eqnarray}}
\newcommand{\eqae}{\end{eqnarray}}
\newcommand{\eqabnon}{\begin{eqnarray*}}
\newcommand{\eqaenon}{\end{eqnarray*}}

\newcommand{\seqb}{\begin{subequations}}
\newcommand{\seqe}{\end{subequations}}

\newcommand{\eref}[1]{\eqref{#1}} 

\newcommand{\defeq}{:=}

\newcommand{\pd}[2]{\frac{\partial #1}{\partial #2}}

\newcommand{\Fds}{F_{\rm ds}}
\newcommand{\Sds}{S_{\rm ds}}
\newcommand{\Tds}{T_{\rm ds}}
\newcommand{\Ads}{A_{\rm ds}}
\newcommand{\sigmads}{\sigma_{\rm ds}}
\newcommand{\Eds}{E_{\rm ds}}

\newtheorem{key}{Key point}
\newtheorem{ass}{Assumption}
\newtheorem{hypo}{Working Hypothesis}

\begin{document}

\markboth{Hiromi Saida}{de~Sitter thermodynamics in the canonical ensemble} 

\title{de~Sitter thermodynamics in the canonical ensemble}

\author{Hiromi \textsc{Saida}\footnote{saida@daido-it.ac.jp}}

\inst{Department of Physics, Daido University, Minami-ku, Nagoya 457--8530, Japan}

\abst{
The existing thermodynamics of the cosmological horizon in de~Sitter spacetime is established in the micro-canonical ensemble, while the thermodynamics of black hole horizons is established in the canonical ensemble. 
Generally in the ordinary thermodynamics and statistical mechanics, both of the micro-canonical and canonical ensembles yield the same equation of state for any thermodynamic system. 
This implies the existence of a formulation of de~Sitter thermodynamics based on the canonical ensemble. 
This paper reproduces the de~Sitter thermodynamics in the canonical ensemble. 
The procedure is as follows: 
We put a spherical wall at the center of de~Sitter spacetime, which has negligible mass and perfectly reflects the Hawking radiation coming from the cosmological horizon. 
Then the region enclosed by the wall and horizon settles down to a thermal equilibrium state, for which the Euclidean action is evaluated and the partition function is obtained. 
The integration constant (subtraction term) of Euclidean action is determined so as to reproduce the equation of state (e.g. entropy-area law) verified already in the micro-canonical ensemble. 
Our de~Sitter canonical ensemble is well-defined in the sense that it preserves the ``thermodynamic consistency'', which means that the state variables satisfy not only the four laws of thermodynamics but also the appropriate differential relations of state variables with thermodynamic functions; e.g. partial derivatives of the free energy yield the entropy, pressure, and so on. 
The special role of cosmological constant in de~Sitter thermodynamics is also revealed. 
}

\PTPindex{451, 454}

\maketitle

\section{Introduction}
\label{sec:intro}

Black hole thermodynamics is well established by the Euclidean action method~\cite{ref:bht,ref:euclidean.sch,ref:euclidean.others}.
Euclidean action method is regarded as one technique to obtain the partition function of canonical ensemble of quantum gravity~\cite{ref:euclidean}. 
(See Appendix~\ref{app:action} for a brief summary of this method.) 
In statistical mechanical sense, the canonical ensemble is suitable for the system under the contact with a heat bath. 
Indeed, as reviewed in next section, thermodynamically consistent formulation of canonical ensemble for black holes has been established by York, where the black hole is in a cavity surrounded by a heat bath~\cite{ref:euclidean.sch} (see Fig.\ref{fig:1} shown in Sec.\ref{sec:ass}). 
The region enclosed by the surface of heat bath and the black hole horizon settles down to a thermal equilibrium state and the Euclidean action method is applied to that region.

On the other hand, thermodynamics of cosmological event horizon (CEH) in de~Sitter spacetime can be formulated for the region bounded by the CEH solely (the region I in Fig.\ref{fig:2} shown in Sec.\ref{sec:ass}), which is filled with the Hawking radiation of CEH and settles down to a thermal equilibrium state without introducing a heat bath. 
This region can be regarded as an isolated thermodynamic system without the contact with heat bath. 
This implies that thermodynamics of CEH can be established in the micro-canonical ensemble. 
Indeed, Hawking and Ross~\cite{ref:dst.micro} have proposed that, when the thermal equilibrium system of a horizon is isolated and has no contact with heat bath, the Euclidean action $I_E^{\rm (micro)}$ of such system can be interpreted as the number of micro-states $W$ or the density of micro-states~\cite{ref:micro} of underlying quantum gravity, which satisfies $W = e^{I_E^{\rm (micro)}}$. 
Then, since $I_E^{\rm (micro)} = \pi r_c^2$ for the isolated system of de~Sitter CEH of radius $r_c$, the entropy-area law can be obtained by the Boltzmann's relation, $\Sds \defeq \ln W = \pi r_c^2$, where $\Sds$ is the entropy of CEH. 
The entropy-area law for CEH, which is the equation of state of CEH, has already been verified in the micro-canonical ensemble.

In the ordinary thermodynamics and statistical mechanics, both the micro-canonical and canonical ensembles yield the same equation of state of any thermodynamic system~\cite{ref:micro,ref:sm}. 
Then it is expected that we can formulate thermodynamics of CEH in the canonical ensemble, which should satisfy the entropy-area law. 
This paper aims to reproduce de~Sitter thermodynamics in the canonical ensemble. 
We will introduce, in next section, a perfectly reflecting spherical wall whose radius is smaller than that of CEH. 
The wall reflects the Hawking radiation coming from CEH, and plays the role of heat bath. 
Then the region enclosed by the wall and CEH settles down to a thermal equilibrium state. 
The Euclidean action of that region is regarded as the partition function, where the integration constant (the so-called subtraction term) of the action should be determined to reproduce the entropy-area law which is the equation of state verified already in the micro-canonical ensemble.

On the other hand, if one wants to regard a theoretical framework as a ``thermodynamics'', the framework must be \emph{thermodynamically consistent}. 
The ``thermodynamic consistency'' means that the state variables satisfy not only the four laws of thermodynamics but also the appropriate differential relations of state variables with thermodynamic functions. 
Thermodynamic functions are the state variables related to the free energy via the Legendre transformation, for example, the internal energy and enthalpy. 
Then a derivative of free energy $F_o$ must yield the entropy $S_o \equiv -\partial F_o/\partial T_o$, where $T_o$ is the temperature. 
There are some other similar differential relations in thermodynamics. 
Those differential relations are the ones required in the ``thermodynamic consistency'', and necessary to understand thermodynamic properties of the system under consideration, for example, phase transition, thermodynamic stability and equations of states.

To ensure the thermodynamic consistency of state variables of de~Sitter CEH, we refer to Schwarzschild thermodynamics formulated by York~\cite{ref:euclidean.sch}, which has precisely established the thermodynamic consistency of the canonical ensemble. 
For example, the work term in the first law and the notion of \emph{extensive} and \emph{intensive} variables are introduced into our de~Sitter thermodynamics with referring to York~\cite{ref:euclidean.sch}. 
Then we can show that the thermodynamic consistency holds for our de~Sitter canonical ensemble.

Here let us comment on the difference of Euclidean action method from ordinary statistical mechanics. 
In the Euclidean action method (see Appendix~\ref{app:action}), one can obtain the partition function of canonical ensemble without explicitly referring to the micro-states of the system under consideration. 
Therefore, although the thermodynamic formulation of black hole is established by that method, the micro-states responsible for the black hole entropy still remain unknown at present. 
Exactly speaking, the black hole thermodynamics at present is still a conjecture, because its micro-states remain unknown. 
Also the micro-states of CEH are still unknown at present. 
This paper does not aim to reveal the micro-states of CEH, but aims to construct the thermodynamically consistent canonical ensemble of CEH in de~Sitter spacetime.

This paper is organized as follows: 
In Sec.\ref{sec:ass}, the basic assumptions of de~Sitter canonical ensemble are introduced with referring to Schwarzschild canonical ensemble formulated by York, where not only the notion of extensive and intensive variables but also the construction of thermal equilibrium system of CEH under the contact with a heat hath are explained. 
Furthermore, a special role of cosmological constant in de~Sitter thermodynamics is also clarified in Sec.\ref{sec:ass}. 
Then Sec.\ref{sec:action} calculates the Euclidean action, where how to determine the integration constant of the action is also discussed. 
Sec.\ref{sec:dst} is devoted to the derivation of important state variables and the exhibition of thermodynamic consistency. 
Sec.\ref{sec:sd} discusses the relation between our Euclidean action and the micro-canonical ensemble proposed already by Hawking and Ross, and also the relation between the cosmological constant and internal energy. 
Appendix~\ref{app:action} reviews the Euclidean action method and shows a clear definition of temperature under the contact with a heat bath. 
Appendix~\ref{app:surface} is for a supplemental explanation about the ``pressure'' introduced in Sec.\ref{sec:dst}, which is not clearly noted in York's paper.

The Planck unit, $c=G=\hbar=k_B=1$, is used throughout.

\section{Basic assumptions to make the Euclidean action work well}
\label{sec:ass}

\subsection{Preliminary}
\label{sec:ass.preliminary}

To make the aim of this section obvious, let us recall the relation between thermodynamics and statistical mechanics~\cite{ref:sm}. 
In statistical mechanics, the partition function can not be expressed as a ``function of state variables'' unless the appropriate state variables, on which the partition function depends, are specified \emph{a priori}. 
To understand this, consider for example an ordinary gas in a spherical container of radius $R$, in which the number of constituent particles is $N$, the mass of one particle is $m$ and the mean velocity of particles is $v$. 
The ordinary statistical mechanics, without the help of thermodynamics, yields the partition function $Z_{\rm gas} = Z_{\rm gas}(R, N, m, v)$ as simply a function of ``parameters'', $R$, $N$, $m$ and $v$. 
Statistical mechanics, solely, can not determine what combinations of those parameters behave as state variables. 
To determine it, the first law of thermodynamics is necessary. 
(Note that the notion of \emph{heat} in the first law is established by purely the argument in thermodynamics, not in statistical mechanics.) 
Comparing the differential of partition function with the first law results in the identification of partition function with the free energy divided by temperature. 
Then, since the free energy of ordinary gases is a function of the temperature and volume due to the ``thermodynamic'' argument, the partition function $Z_{\rm gas}(R, N, m, v)$ should be rearranged to be a function of temperature and volume $Z_{\rm gas}(V, T)$, where $V = (4 \pi/3) R^3$ and $T = m v^2$ for ideal gases due to the law of equipartition of energy~\footnote{
When the number of particles $N$ changes by, for example, a chemical reaction and an exchange of particles with environment, $N$ is also the state variable on which the free energy depends.
}. 
(The dependence on $N$ is, for example, $Z_{\rm gas} \propto N$ for ideal gases.)

It should also be emphasized that the reason why the temperature and volume are regarded as the state variables of the gas is that they are consistent with the four laws of ``thermodynamics'' and have the appropriate properties as state variable. 
The appropriate properties are that the state variables are macroscopically measurable, the state variables are distinguished into two categories, \emph{intensive} variables and \emph{extensive} variables, and the extensive variables are additive. 
Those properties of state variables are specified by purely the argument in thermodynamics, not in statistical mechanics. 
Therefore, from the above, it is recognized that statistical mechanics can not yield the partition function as a ``function of appropriate state variables'' without the help of thermodynamics which specifies the appropriate state variables for the partition function.

Turn our discussion to the Euclidean action method for black hole and de~Sitter spacetimes. 
Since the Euclidean action method is the technique to obtain the ``partition function'' of the spacetime under consideration (see Appendix~\ref{app:action}), it is necessary to specify the state variables before calculating the Euclidean action. 
In this section, we review the canonical ensemble for Schwarzschild spacetime constructed by York~\cite{ref:euclidean.sch}. 
Then, from the Schwarzschild canonical ensemble, we can learn the appropriate state variables for the partition function in de~Sitter canonical ensemble, which will be summarized later as the basic assumptions of de~Sitter canonical ensemble. 
Furthermore, the special role of cosmological constant is also clarified. 
The calculation of Euclidean action is carried out not in this section but in next section.

\subsection{Schwarzschild canonical ensemble}
\label{sec:ass.sch}

\begin{figure}[t]
 \begin{center}
 \includegraphics[height=30mm]{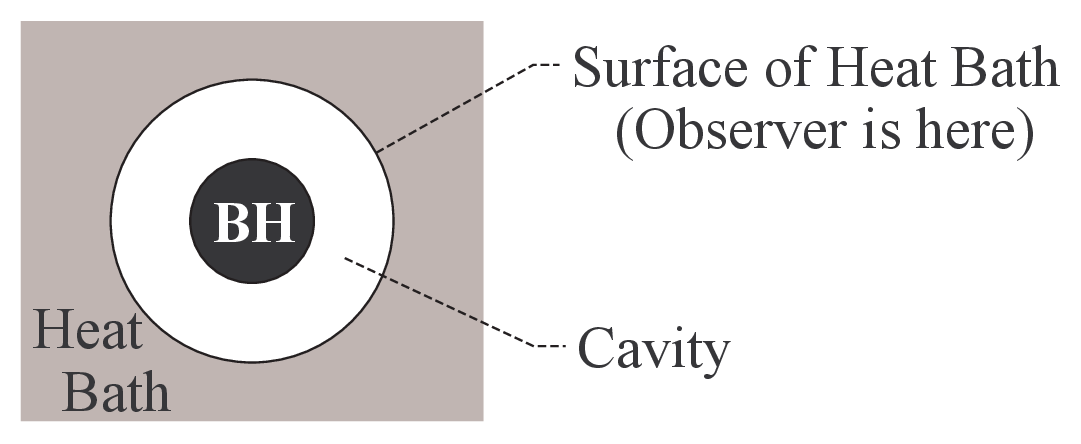}
 \end{center}
\caption{Schematic image of thermal equilibrium of black hole with heat bath. This is described in the canonical ensemble. State variables of black hole are defined at the surface of heat bath. With those state variables, the consistent thermodynamic formulation is realized using the Euclidean action method~\cite{ref:euclidean.sch}.}
\label{fig:1}
\end{figure}

In the canonical ensemble for Schwarzschild thermodynamics, there are three key points which imply the basis of the canonical ensemble for de~Sitter thermodynamics. 
The first one is the zeroth law which describes the existence and construction of thermal equilibrium states: 
\begin{key}[Zeroth law of black hole] 
Place a black hole in a spherical cavity as shown in Fig.\ref{fig:1} and also the observer at the surface of the heat bath. 
Through the Hawking radiation by black hole and the black body radiation by heat bath, the black hole interacts with the heat bath. 
Then, by appropriately adjusting the temperature of heat bath, the combined system of black hole and heat bath settles down to a thermal equilibrium state.
\end{key}

The equilibrium state of black hole under the contact with heat bath is described in the canonical ensemble. 
And the equilibrium state variables of black hole are defined by the quantities measured at the surface of heat bath where the observer is. 
Then the ``thermodynamically consistent'' Schwarzschild canonical ensemble is constructed as follows.

The second key point is the difference of black hole thermodynamics from thermodynamics of ordinary laboratory systems: 
\begin{key}[Peculiar scaling law of black hole] 
Extensive and intensive state variables of black hole show a peculiar scaling law: 
When a length size $L$ (e.g. event horizon radius) is scaled as $L \to \lambda\,L$ with an arbitrary scaling rate $\lambda \, (>0)$, then the extensive variables $X$ (e.g. entropy) and intensive variables $Y$ (e.g. temperature) are scaled as $X \to \lambda^2\,X$ and $Y \to \lambda^{-1}\,Y$, while the thermodynamic functions $\Phi$ (e.g. free energy) are scaled as $\Phi \to \lambda\,\Phi$. 
This implies that, because the system size is one of the extensive variables, the thermodynamic system size of equilibrium system constructed in the key point~1 should have the areal dimension. 
Indeed the surface area of heat bath, $4 \pi r_w^2$, behaves as the consistent extensive variable of the system size, where $r_w$ is the radius of the surface of heat bath. 
\end{key}

Here recall that, in thermodynamics of ordinary laboratory systems, the intensive variables remain un-scaled under the scaling of system size, the extensive variables scales as the volume, and the thermodynamic functions are the members of extensive variables. 
However, as noted in the key point~2, the black hole thermodynamics has the peculiar scaling law of state variables. 
Although the scaling law differs from that in thermodynamics of ordinary laboratory systems, the peculiar scaling law in black hole thermodynamics retains the thermodynamic consistency as noted in the next key point.

The third key point is the similarity of black hole thermodynamics with thermodynamics of ordinary laboratory systems:
\begin{key}[Euclidean action method and thermodynamic consistency] 
The free energy $F_{\rm BH}$ of black hole is yielded by the Euclidean action method, where the integration constant (the so-called subtraction term) of the action integral is determined with referring to flat spacetime. 
The action integral is evaluated in the region, $2 M < r < r_w$, which is in thermal equilibrium as noted in the key point~1. 
(Here $M$ is the mass parameter.) 
Then, as for the ordinary thermodynamics, this free energy is expressed as a function of two independent state variables, temperature and system size;
\eqb
 F_{\rm BH}(T_{\rm BH} , 4 \pi r_w^2) \,,
\eqe
where the intensive variable $T_{\rm BH} \defeq \left(8 \pi M \sqrt{1-2 M/r_w}\right)^{-1}$ is the Hawking temperature measured by the observer at $r_w$, and the factor $\sqrt{1-2 M/r_w}$ is the so-called Tolman factor~\cite{ref:tolman} which expresses the gravitational redshift affecting the Hawking radiation propagating from the black hole horizon to the observer. 
This Hawking temperature is obtained by Eq.\eref{eq:app.action.T} of Appendix~\ref{app:action}. 
In order to let $T_{\rm BH}$ and $4 \pi r_w^2$ be independent variables in $F_{\rm BH}$, the mass parameter $M$ and the heat bath radius $r_w$ are regarded as two independent variables. 
Then the thermodynamic consistency holds as follows: 
The entropy $S_{\rm BH}$ and the ``surface pressure'' $\sigma_{\rm BH}$ are defined by
\eqb
\label{eq:intro.S.sigma_BH}
 S_{\rm BH} \defeq -\pd{F_{\rm BH}(T_{\rm BH} , 4 \pi r_w^2)}{T_{\rm BH}}
                     = \frac{A_{\rm BH}}{4} \quad,\quad
 \sigma_{\rm BH} \defeq -\pd{F_{\rm BH}(T_{\rm BH} , 4 \pi r_w^2)}{ (4 \pi r_w^2)} \,,
\eqe
where $\sigma_{\rm BH}$ has the dimension of force par unit area because the system size $4 \pi r_w^2$ has the dimension of area. 
See Appendix~\ref{app:surface} for a detail explanation of thermodynamic meaning of $\sigma_{\rm BH}$. 
(The temperature of heat bath should be adjusted to be $T_{\rm BH}$ in the key point~1.) 
These differential relations among the free energy, entropy and surface pressure are the same with those obtained in thermodynamics of ordinary laboratory systems. 
Furthermore, as for the ordinary thermodynamics, the internal energy and the other thermodynamic functions are defined by the Legendre transformation of the free energy; for example the internal energy $E_{\rm BH}$ is
\eqb
 E_{\rm BH}(S_{\rm BH},4 \pi r_w^2) \defeq F_{\rm BH} + T_{\rm BH}\,S_{\rm BH} \,.
\eqe
The enthalpy, Gibbs energy and so on are also defined by the Legendre transformation. 
Then the differential relations among those thermodynamic functions and the other state variables also hold, for example $T_{\rm BH} \equiv \partial E_{\rm BH}/\partial S_{\rm BH}$. 
Furthermore, with the state variables obtained above, we can check that the first, second and third laws of thermodynamics hold for black holes. 
\end{key}

The above three key points hold also for the other single-horizon black hole spacetimes, and those black hole thermodynamics has already been established~\cite{ref:euclidean.sch,ref:euclidean.others}.

Here let us remark about the heat bath introduced in the key point~1. 
In York's consistent black hole thermodynamics~\cite{ref:euclidean.sch}, the heat bath is essential to establish the thermodynamic consistency in the canonical ensemble as explained below: 
Generally in thermodynamics, as noted in the key point~3, thermodynamic functions are defined as a function of two independent state variables. 
Especially the free energy should be expressed as a function of the temperature and the extensive state variable which represents the system size. 
This thermodynamic requirement is satisfied by introducing the heat bath, which gives us two independent variables; the mass parameter $M$ and the radius of heat bath $r_w$. 
These two independent variables makes it possible to define the temperature $T_{\rm BH}$ and the surface area $4 \pi r_w^2$ as the two independent state variables in free energy $F_{\rm BH}(T_{\rm BH},4\pi r_w^2)$. 
Therefore the heat bath is necessary to establish manifestly the thermodynamic consistency.

Concerning the heat bath, let us make another comment here. 
It is possible to take the limit $r_w \to \infty$ after constructing the consistent black hole thermodynamics with the heat bath of finite $r_w$.
Here one may think that the limit $r_w \to \infty$ corresponds to the micro-canonical ensemble, since state variables are expressed as functions of only one parameter $M$ and the heat bath seems to disappear (run away to infinitely distant region). 
However it should be emphasized that, generally in statistical mechanics, the micro-canonical ensemble is not some limiting case of the canonical ensemble. 
Therefore the limit $r_w\to\infty$ does not mean to consider the micro-canonical ensemble (the system without heat bath), but it is just the large limit of the cavity size in the canonical ensemble. 
Hence the black hole thermodynamics have been established in the canonical ensemble.

\subsection{Basic assumptions of de~Sitter canonical ensemble}
\label{sec:ass.basic}

Let us proceed to the introduction of basic assumptions of de~Sitter canonical ensemble. 
Those assumptions give us the way to construct thermal equilibrium state under the contact with heat bath, and specifies the appropriate state variables for the partition function. 
Furthermore, to make our discussion exact logically, the use of Euclidean action method is also listed up as one assumption in this paper.~\footnote{In this paper, we take the standpoint that the Euclidean action method is simply one (promising) way of obtaining the partition function of spacetimes. At present, since we do not know the complete quantum gravity theory, the use of Euclidean action is understood as one assumption.}

\begin{figure}[t]
 \begin{center}
 \includegraphics[height=30mm]{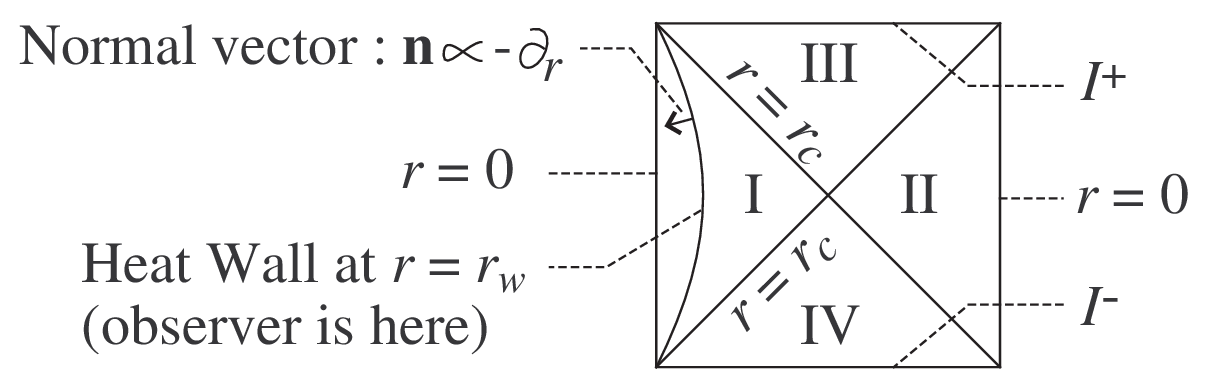}
 \end{center}
\caption{Penrose diagram of de~Sitter spacetime. $I^{\pm}$ is the future/past null infinity. $r_c$ is the CEH radius. The heat wall of radius $r_w$ is placed at the center of the region surrounded by CEH. Our observer is at the wall. This wall reflects perfectly the Hawking radiation, and the observer sees that the region enclosed by the wall and CEH is in a thermal equilibrium state.}
\label{fig:2}
\end{figure}

The cosmological event horizon (CEH) in de~Sitter spacetime is a spherically symmetric null hyper-surface defined as the boundary of causal past of the observer's world line~\cite{ref:temperature}. 
The observer detects the Hawking radiation of thermal spectrum emitted by the CEH~\cite{ref:temperature}. 
This implies that the observer can regard the CEH as an object in thermal equilibrium. 
It is expected that a thermodynamically consistent canonical ensemble of CEH can be constructed in the way similar to the Schwarzschild canonical ensemble. 
Then, referring to the key point~1 of Schwarzschild thermodynamics, the first basic assumption of de~Sitter canonical ensemble is the zeroth law:
\begin{ass}[Zeroth law of CEH]
Place a spherically symmetric thin wall of radius $r_w$ at the center of the region surrounded by the CEH as shown in Fig.\ref{fig:2}. 
The wall is smaller than CEH, $r_w < r_c$, where $r_c$ is the CEH radius. 
The mass energy of this wall is negligible, and the geometry of the region $r_w < r < r_c$ is of de~Sitter spacetime. 
Let the wall reflect perfectly the Hawking radiation, and we call this perfectly reflecting wall the ``heat wall'' hereafter. 
We put our observer at the heat wall. 
Then this observer sees that the region enclosed by the heat wall and CEH settles down to a thermal equilibrium state. 
The equilibrium state variables of CEH are defined by the quantity measured at the heat wall where the observer is.
\end{ass}

Since this equilibrium state of CEH has the contact with the heat wall, we expect that de~Sitter thermodynamics is obtained in the canonical ensemble. 
However before calculating the partition function, as explained in Sec.\ref{sec:ass.preliminary}, we have to specify the appropriate state variables for the partition function, the temperature and extensive variable of system size. 
To do so, it is necessary to clarify the notion of extensivity and intensivity, which is the significant scaling property of state variables under the scaling of system size. 
The scaling behavior provides a guideline to specify the state variable of system size, and also provides the basis to adopt the temperature defined by Eq.\eref{eq:app.action.T} of Appendix~\ref{app:action} as an intensive state variable (see Sec.\ref{sec:dst.scaling}).

Concerning the extensivity and intensivity, note that, in general, thermodynamics seems to be a very universal formalism which can be applied to any system if it is in thermal equilibrium. 
This implies that the scaling behavior of state variables shown in the key point~2 of Schwarzschild thermodynamics is common to any horizon system. 
Then the second assumption of de~Sitter canonical ensemble is as follows:
\begin{ass}[Scaling law and system size of CEH]
State variables of CEH are classified into three categories, extensive variables, intensive variables and thermodynamic functions, and those state variables satisfy the same scaling law as explained in the key point~2: 
When a length size $L$ (e.g. CEH radius) is scaled as $L \to \lambda\,L$ with an arbitrary scaling rate $\lambda \, (>0)$, then the extensive variables $X$ (e.g. system size) and intensive variables $Y$ (e.g. temperature) are scaled as $X \to \lambda^2\,X$ and $Y \to \lambda^{-1}\,Y$, while the thermodynamic functions $\Phi$ (e.g. free energy) are scaled as $\Phi \to \lambda\,\Phi$. 
This implies that the thermodynamic system size of the equilibrium system constructed in assumption~1 should have the areal dimension, and the surface area of heat wall, $\Ads \defeq 4 \pi r_w^2$, behaves as the extensive variable of system size. 
(Indeed, it will be shown later that thermodynamically consistent formulation of de~Sitter canonical ensemble can be established with this $\Ads$.) 
\end{ass}

The concrete functional form of free energy can not be determined by only the assumption~2. 
But, as implied by the key point~3 of Schwarzschild thermodynamics, the Euclidean action method can yield the concrete form of free energy. 
Hence the third assumption of de~Sitter canonical ensemble is the declaration of using the Euclidean action method:
\begin{ass}[Euclidean action and State variables of CEH] 
Euclidean action $I_E$ of a thermal equilibrium state of CEH yields the partition function of  canonical ensemble by Eq.\eref{eq:app.action.Zcl} of Appendix~\ref{app:action}, where the integration in $I_E$ is calculated over the region enclosed by the heat wall and CEH ($r_w < r < r_c$). 
And the free energy $\Fds(\Tds,\Ads)$ of CEH is defined by Eq.\eref{eq:app.action.F}, where $\Tds$ is the temperature of CEH determined by Eq.\eref{eq:app.action.T} and $\Ads$ is the area of heat wall discussed in the assumption~2. 
Once $\Fds$ is determined, any state variable of CEH is defined from $\Fds$ in the same way with the state variable in thermodynamics of ordinary laboratory systems. 
For example, CEH entropy $\Sds$ is defined by $\Sds \defeq -\partial \Fds/\partial \Tds$.
\end{ass}

The definition of Euclidean action $I_E$ is shown in Eq.\eref{eq:app.action.IE.curved} of Appendix~\ref{app:action}, where the Lorentzian action $I_L$ is used. 
The Lorentzian Einstein-Hilbert action $I_L$ is
\eqb
 I_L \defeq
  \frac{1}{16 \pi}\int_{\mathcal{M}} dx^4\,\sqrt{-g}\,\left( \mathcal{R} - 2\,\Lambda \right)
 + \frac{1}{8\,\pi}\int_{\partial \mathcal{M}} dx^3\,\sqrt{h}\,K
 + I_{L {\rm sub}} \, ,
\label{eq:ass.IL}
\eqe
where $\mathcal{M}$ is the region enclosed by the heat wall and CEH in de~Sitter spacetime ($r_w < r < r_c$), $\mathcal{R}$ is the Ricci scalar, $\Lambda$ is the cosmological constant, $g$ is the determinant of the metric, $h$ and $K$ in the second term are respectively the determinant of first fundamental form (induced metric) and the trace of second fundamental form (extrinsic curvature) of the boundary ${\partial \mathcal{M}}$ (which is the world volume of heat wall), and $I_{L {\rm sub}}$ is the integration constant of $I_L$ which is sometimes called the subtraction term. 
The second term $\int_{\partial M}$ in Eq.\eref{eq:ass.IL} is required to eliminate the second derivatives of metric from the action~\cite{ref:action}. 
The third term $I_{L {\rm sub}}$ does not contribute to the Einstein equation obtained by $\delta I_L = 0$, and is specified in the working hypothesis~2 explained below. 
The Einstein equation for de~Sitter spacetime yields the relation $\mathcal{R} = 4\,\Lambda$. 
Using this $I_L$, the Euclidean action $I_E$ for our thermal system constructed in assumption~1 is calculated in next section.

In Sec.\ref{sec:dst}, we will suggest that the assumptions~1,~2 and~3 construct a candidate of the ``thermodynamically consistent'' de~Sitter canonical ensemble. 
Let us emphasize that, exactly and logically speaking, once these assumptions are adopted, we also have to adopt supplemental working hypotheses as explained in next subsection.

\subsection{Supplemental working hypotheses}

When we require the thermodynamic consistency, the differential relations like Eq.\eref{eq:intro.S.sigma_BH} must be explicitly satisfied. 
And note that the free energy $\Fds$ should be a function of two independent variables $\Tds$ and $\Ads$. 
These require the existence of two independent state variables. 
On the other hand, we have only two parameters $\Lambda$ and $r_w$ in our equilibrium system constructed in assumption~1. 
Hence we have to adopt the following working hypothesis:
\begin{hypo}[Two independent variables]
To construct de~Sitter canonical ensemble in a thermodynamically consistent way, we require that the cosmological constant $\Lambda$ is an independent ``working variable''. 
Then we have two independent variables $\Lambda$ and $r_w$ which can ensure the free energy to be a function of two independent state variables. 
When we regard non-variable $\Lambda$ as the physical situation, it is obtained by the ``constant $\Lambda$ process'' in the ``generalized'' de~Sitter thermodynamics in which $\Lambda$ is regarded as a working variable to ensure the thermodynamic consistency.
\end{hypo}

A verification of this working hypothesis will be discussed in Sec.\ref{sec:sd}. 
This working hypothesis can be combined with the assumption~3 to obtain a ``complete free energy'' as a function of two independent state variables. 
Here, in order to emphasize the role of $\Lambda$ as a working variable, we divide the requirement of ``complete free energy'' into the assumption~3 and the working hypothesis~1. 
The assumption~3 and working hypothesis~1 may be combined to form one assumption.

Here let us comment on this working hypothesis. 
This paper is not the first which requires the variable $\Lambda$. 
For example, in order to consider a thermodynamic formalism for Schwarzschild-de~Sitter black hole, it has been reported that $\Lambda$ has to be regarded as an variable, otherwise thermodynamic consistency is lost~\cite{ref:lambda}. 
The variable $\Lambda$ seems to be a practical idea/assumption to treat CEH in thermodynamic framework.

Under the working hypothesis~1, the Euclidean action is to be expressed as a function of two ``independent'' state variables, temperature and surface area of heat wall (see the assumption~2). 
Of course, also the integration constant $I_{L {\rm sub}}$ in Eq.\eref{eq:ass.IL} should be determined. 
Here recall that, as mentioned in Sec.\ref{sec:intro} (third paragraph), the entropy-area law should be reproduced from our Euclidean action of de~Sitter canonical ensemble, since this law is the equation of state verified already in de~Sitter micro-canonical ensemble~\cite{ref:dst.micro,ref:micro}. 
This requirement gives us the guiding principle to determine $I_{L {\rm sub}}$, which we summarize in the following working hypothesis:
\begin{hypo}[Consistency with the micro-canonical ensemble] 
The entropy-area law, which is the equation of state verified already in the micro-canonical ensemble, should be reproduced in our de~Sitter canonical ensemble. 
In other words, the integration constant $I_{L {\rm sub}}$ in the action $I_L$ should be determined so as to reproduce the entropy-area law. 
To do so, in calculating the Euclidean action of de~Sitter spacetime, we set for the time being,
\eqb
\label{eq:ass.IEsub}
 I_{E {\rm sub}} = \alpha_w \,I^{\rm (flat)}_E \, ,
\eqe
where $I_{E {\rm sub}}$ is the integration constant in de~Sitter Euclidean action, $I^{\rm (flat)}_E$ is the Euclidean action of flat spacetime ($\Lambda = 0$) whose concrete form will be shown in Sec.\ref{sec:action}, and $\alpha_w$ is a dimensionless factor. 
In order not to let $I_{E {\rm sub}}$ affect the variational principle in obtaining Einstein equation, the factors $\alpha_w$ and $I^{\rm (flat)}_E$ should be expressed by only the quantity determined at the boundary of Euclidean de~Sitter space (at the heat wall). 
Indeed, as will be shown in Sec.\ref{sec:action}, $I^{\rm (flat)}$ is expressed by only such quantity. 
The concrete form of $\alpha_w$ will be obtained in Sec.\ref{sec:dst.IE}.
\end{hypo}

Let us emphasize that Eq.\eref{eq:ass.IEsub} is just a working hypothesis including the unknown factor $\alpha_w$. 
It seems that, even if one starts calculations of de~Sitter Euclidean action with $I_{E {\rm sub}}$ which is proportional (not to $I^{\rm (flat)}_E$ but) to some other action integral determined by only the boundary of Euclidean space, then the requirement of preserving the entropy-area law results in the same form with our $I_{E {\rm sub}}$ obtained in Sec.\ref{sec:dst.IE}. 
The essence of the working hypothesis~2 is not Eq.\eref{eq:ass.IEsub} but the preservation of entropy-area law which is the equation of state verified already in the micro-canonical ensemble. 
This working hypothesis~2 is based on the statistical mechanical requirement that both the micro-canonical and canonical ensembles yield the same equation of state for any thermodynamic system~\cite{ref:micro,ref:sm}.

\section{Euclidean action for canonical ensemble}
\label{sec:action}

We calculate the Euclidean action in this section. 
But, for the completeness of discussion, let us summarize the Lorentzian and Euclidean de~Sitter metric before calculating the Euclidean action.

\subsection{Euclidean de~Sitter space}
\label{sec:action.space}

The Lorentzian de~Sitter metric in static chart is
\eqb
 ds^2 = -f(r)\,dt^2 + \frac{dr^2}{f(r)} + r^2\,d\Omega^2 \, ,
\eqe
where $d\Omega^2 = d\theta^2 + \sin^2\theta\,d\varphi^2$ is the line element on a unit two-sphere, and
\eqb
 f(r) \defeq 1 - H^2\,r^2 \quad,\quad 3\,H^2 = \Lambda \,.
\eqe
The Penrose diagram is shown in Fig.\ref{fig:2}. 
The static chart with $0 \le r < H^{-1}$ covers the region I or II shown in Fig.\ref{fig:2}. 
The notion of CEH is observer dependent~\cite{ref:temperature}. 
For the observer whose world line is confined in the region I, the radius $r_c$ of CEH is given by $f(r_c) = 0$ ;
\eqb
 r_c = \frac{1}{H} \, .
\eqe
A Killing vector $\xi \propto \partial_t$ becomes null at the CEH. 
This means that the CEH is the Killing horizon of $\xi$. 
Surface gravity $\kappa$ of the Killing horizon, CEH, is defined by the relation, $\nabla_\xi\,\xi = \kappa\,\xi$ at the CEH. 
The value of $\kappa$ depends on the normalization of $\xi$ by definition~\cite{ref:sg}. 
When the Killing vector is normalized as $\xi = \partial_t$, we get
\eqb
 \kappa = H \, .
\label{eq:action.sg}
\eqe
This $\kappa$ is the surface gravity of CEH measured by the observer at the central world line $r = 0$, because $t$ is the proper time of the central observer and the norm $\xi^2 = - f(r)$ is $-1$ at $r = 0$.

The global chart of Lorentzian de~Sitter spacetime is introduced by the coordinate transformation from $(t,r,\theta,\varphi)$ to $(W,X,\theta,\varphi)$,
\eqb
 W-X \defeq e^{H\,(t-r^{\ast})} \quad,\quad W+X \defeq - e^{-H\,(t+r^{\ast})} \, ,
\label{eq:action.dS.trans}
\eqe
where $dr^{\ast} \defeq dr/f(r)$ which means $r^{\ast} = (1/2 H)\,\ln\left|(1 + H\,r)/(1 - H\,r)\right|$. 
This transformation yields
\eqb
\label{eq:action.dS.global}
 ds^2 = \frac{1}{H^2 \left( W^2 - X^2 -1 \right)^2}
        \left[ -dW^2 + dX^2 + \frac{\left( W^2 - X^2 + 1 \right)^2}{4}\,d\Omega^2 \right] \, .
\eqe
The coordinate transformation~\eref{eq:action.dS.trans} implies the range of coordinates, $X < W < -X$ and $ X < 0$. 
By extending it to the range $-\infty < W < \infty$ and $-\infty < X < \infty$, the global chart covers the whole region of maximally extended de~Sitter spacetime, I, II, III and IV shown in Fig.\ref{fig:2}.

Here concerning the relation of the two charts, note that the Penrose diagram shown in Fig.\ref{fig:2} is depicted according to the global chart~\eref{eq:action.dS.global}. 
Under the coordinate transformation~\eref{eq:action.dS.trans}, the direction of timelike Killing vector $\xi \defeq \partial_t$ is future pointing in the region I, and past pointing in the region II. 
Therefore, throughout this paper, we consider the region I in using the static chart~\footnote{If the signature of exponent in the transformation~\eref{eq:action.dS.trans} is opposite, $W - X = \exp[-H (t-r^{\ast})]$ and $W + X = -\exp[H (t+r^{\ast})]$, then the direction of $\partial_t$ becomes past pointing in the region I, and future pointing in the region II.}.

Next we proceed to the construction of Euclidean de~Sitter space. 
As explained in Appendix~\ref{app:action}, we apply the Wick rotations $t \to - i \tau$ in the static chart and $W \to - i w$ in the global chart to obtain the Euclidean de~Sitter space. 
These Wick rotations are equivalent, because the transformation~\eref{eq:action.dS.trans}, $W = e^{-H r^{\ast}}\sinh\left(H t\right)$, implies that the imaginary time $w$ in global chart is defined by $w \defeq e^{-H r^{\ast}}\sin\left(H \tau\right)$, where $\tau$ is the imaginary time in the static chart. 
The Euclidean metric in the static chart is
\eqb
 ds_E^2 = f(r)\,d\tau^2 + \frac{dr^2}{f(r)} + r^2\,d\Omega^2 \, .
\label{eq:action.dSE.static} \\
\eqe
The Euclidean metric in the global chart is
\eqb
 ds_E^2 = \frac{1}{H^2 \left( w^2 + X^2 + 1 \right)^2}
          \left[ dw^2 + dX^2 + \frac{\left( w^2 + X^2 - 1 \right)^2}{4}\,d\Omega^2 \right] \, .
\label{eq:action.dSE.global}
\eqe
About the global chart, we get from the coordinate transformation~\eref{eq:action.dS.trans},
\eqb
 w^2 + X^2 = \frac{1 - H\,r}{1 + H\,r} \, .
\eqe
Because of $w^2 + X^2 \ge 0$, the Euclidean de~Sitter space corresponds to the region I (or~II), $0<r<r_c$, in the Lorentzian de~Sitter spacetime. 
Because the Lorentzian de~Sitter spacetime is regular at $r = 0$ and $r = r_c$, the Euclidean de~Sitter space is also regular at those points. 
To examine the regularity of Euclidean space, we make use of the metric in static chart~\eref{eq:action.dSE.static}. 
The regularity at $r = 0$ is rather obvious, since the metric near $r = 0$ is flat and regular, $ds_E^2 \simeq d\tau^2 + dr^2 + r^2\,d\Omega^2$.

To examine the regularity of Euclidean space at $r = r_c$, let us define a coordinate $y$
 and a function $b(y)$ by
\eqb
 y^2 \defeq r_c - r \quad,\quad
 b(y) \defeq \sqrt{f(r_c - y^2)} \, .
\eqe
We get $b'(y) \defeq d b(y)/dy = 2\,H^2\,r\,y/b(y)$ which denotes
\eqb
 \lim_{y \to 0}b'(y) = 2\,H\,\lim_{y\to 0}\frac{y}{b(y)}
                     = 2\,H\,\frac{1}{\displaystyle \lim_{y \to 0}b'(y)} \, .
\eqe
This means $b'(0) = \sqrt{2\,H}$, and near the CEH, $f \simeq \left[\, b(0) + b'(0)\,y \,\right]^2 = 2\,H\,y^2$. 
Therefore the Euclidean metric near the CEH is
\eqb
 ds_E^2 \simeq
 \frac{2}{H}\,\left[\, y^2\,d(H \tau)^2 + dy^2 \,\right] + \frac{1}{H^2}\,d\Omega^2 \, .
\eqe
It is obvious that the Euclidean de~Sitter space is regular at CEH if the imaginary time has the period $\beta$ defined by
\eqb
 0 \le \tau < \beta \defeq \frac{2\,\pi}{H} \, .
\label{eq:action.period}
\eqe
Throughout our discussion, the imaginary time $\tau$ has the period $\beta$.

\subsection{Euclidean action for ``thermodynamically consistent'' de~Sitter thermodynamics}
\label{sec:action.IE}

Let us proceed to the calculation of Euclidean action $I_E$ defined by Eq.\eref{eq:app.action.IE.curved}. 
To obtain $I_E$, we should specify the Lorentzian action $I_L$ given in Eq.\eref{eq:ass.IL}. 
The integral region ${\mathcal M}$ in $I_L$ is the Lorentzian region which forms the thermal equilibrium state constructed in the assumption~1. 
Therefore ${\mathcal M}$ is expressed by $r_w < r < r_c$, and its boundary $\partial{\mathcal M}$ is at the heat wall, $r = r_w$. 
There is another boundary at $r = r_c$ in the Lorentzian region ${\mathcal M}$. 
However we do not need to consider it, because, as shown at the end of previous subsection, the points at $r = r_c$ in \emph{Euclidean} space do not form a boundary but are the regular points when the imaginary time $\tau$ has the period~\eref{eq:action.period}. 
Then the first fundamental form $h_{i j}$ ($i, j = 0, 2, 3$) of $\partial{\mathcal M}$ in the static chart (in Lorentzian de~Sitter spacetime) is
\eqb
 \left. ds^2 \right|_{r = r_w} = h_{i j}\, dx^i \, dx^j
 = - f_w\,dt^2 + r_w^2\,d\Omega^2 \, ,
\eqe
where
\eqb
 f_w \defeq f(r_w) = 1 - H^2\,r_w^2 \, .
\label{eq:action.fw}
\eqe
Here, since ${\mathcal M}$ is the region enclosed by the heat wall and CEH, the direction of unit normal vector $\mathbf{n}$ to $\partial{\mathcal M}$ is pointing towards the ``center'' $r=0$, which means $\mathbf{n} \propto - \partial_r$ (see Fig.\ref{fig:2}). 
Then the second fundamental form of $\partial{\mathcal M}$ in the static chart is
\eqb
 K_{i j} =
 - \sqrt{f_w}\, {\rm diag.}\left[\, H^2\,r_w \,,\, r_w \,,\, r_w\,\sin^2\theta \, \right] \, ,
\label{eq:action.Kij}
\eqe
where diag. means the diagonal matrix form.

On the other hand, the second fundamental form $K_{i j}^{\rm (flat)}$ of a spherically symmetric timelike hyper-surface $r = r_w$ in Minkowski spacetime is obtained by setting $H=0$ in Eq.\eref{eq:action.Kij},
\eqb
 K_{i j}^{\rm (flat)} =
 - {\rm diag.}\left[\, 0 \,,\, r_w \,,\, r_w\,\sin^2\theta \, \right] \, ,
\eqe
where the normal vector $\mathbf{n}^{\rm (flat)}$ to this surface is set to be $\mathbf{n}^{\rm (flat)} \propto - \partial_r$ as that in $K_{i j}$. 
For Minkowski spacetime $\mathcal{R} = 0$ and $\Lambda=0$, and its Lorentzian action $I^{\rm (flat)}$ is expressed by only the surface term in Eq.\eref{eq:ass.IL},
\eqb
 I^{\rm (flat)} = \frac{1}{8 \pi}\int_{\partial \mathcal{M}} dx^3\,\sqrt{h}\,K^{\rm (flat)} \,.
\eqe
In applying this $I^{\rm (flat)}$ to Eq.\eref{eq:ass.IEsub} of working hypothesis~2, the integral element $\sqrt{h}$ in $I^{\rm (flat)}$ should be that of de~Sitter spacetime, because the background spacetime on which the integral is calculated is the de~Sitter spacetime.

From the above, applying the Wick rotation $t \to -i \tau$ to the Lorentzian action $I_L$ of de~Sitter spacetime, we obtain the Euclidean action $I_E$ via Eqs.\eref{eq:app.action.IE.curved} and~\eref{eq:ass.IEsub},
\eqab
I_E
 &=& \frac{3\,H^2}{8\,\pi} \int_{\mathcal M_E}dx_E^4\,\sqrt{g_E}
   + \frac{1}{8\,\pi}\int_{\partial\mathcal M} dx_E^3 \,\sqrt{h_E}\,
     \left(\,K_E + \alpha_w\,K_E^{\rm (flat)}\,\right) \nonumber \\
 &=& \frac{\pi}{H^2}\,\left(\, 1 - 2\,H\,r_w\,f_w - 2\,\alpha_w\,H\,r_w\,\sqrt{f_w}\,\right) \, ,
\label{eq:action.IE}
\eqae
where the relation for de~Sitter spacetime $\mathcal{R} = 4\,\Lambda = 12\,H^2$ is used in the first equality, $Q_E$ is the quantity $Q$ evaluated on the Euclidean de~Sitter space, and ${\mathcal M_E}$ is the Euclidean region expressed by $0 \le \tau < \beta$ , $r_w \le r \le r_c$ , $0 \le \theta \le \pi$ and $0 \le \varphi < 2\,\pi$. 
This $I_E$ corresponds to $I_E[g_{E\,cl}]$ in Eq.\eref{eq:app.action.F} of Appendix~\ref{app:action}, which yields the partition function for the thermal equilibrium of spacetime with neglecting quantum fluctuations.

A comment on the limit $r_w \to 0$ of this $I_E$ will be given in Sec.\ref{sec:sd} after specifying the form of $\alpha_w$ in next section.

\section{de Sitter thermodynamics in the canonical ensemble}
\label{sec:dst}

In this section, principal state variables of CEH are calculated successively, and the ``thermodynamic consistency'' is also shown explicitly.

\subsection{Temperature}
\label{sec:dst.T}

As explained in Eq.\eref{eq:app.action.T} of Appendix~\ref{app:action}, the temperature $\Tds$ of CEH is defined by the integral in imaginary time direction at the heat wall,
\eqb
 \Tds \defeq \left[\, \int_0^{\beta}\,\sqrt{f_w}\,d\tau \,\right]^{-1}
      = \frac{H}{2 \pi \sqrt{f_w}} \, ,
\label{eq:dst.T}
\eqe
where $\beta$ is defined in Eq.\eref{eq:action.period} and $f_w$ is in Eq.\eref{eq:action.fw}. 
Note that, as implied by Eq.\eref{eq:action.sg}, $H/2\pi$ coincides with the Hawking temperature measured at $r=0$~\cite{ref:temperature}, and the factor $\sqrt{f_w}$ in Eq.\eref{eq:dst.T} is the Tolman factor which expresses the gravitational redshift affecting the Hawking radiation propagating from CEH to the observer~\cite{ref:tolman}. 
This $\Tds$ is the temperature measured by the observer at heat wall.

\subsection{Surface area and homothetic variation of the system}
\label{sec:dst.A}

In the ordinary laboratory systems, the size of the system under consideration is its volume. 
For de~Sitter spacetime, the volume can not be defined uniquely, since the choice of spatial slice in the region ${\mathcal M}$ of $r_w < r < r_c$ is not determined in a natural way. 
However, an area can be uniquely and naturally assigned to our system. 
It is the surface area $\Ads$ of the heat wall.

The above discussion supports the assumption~2 in which $\Ads$ is regarded as the state variable. 
Therefore, we adopt $\Ads$ as the extensive state variable of system size,
\eqb
 \Ads \defeq 4\,\pi\,r_w^2 \, .
\label{eq:dst.A}
\eqe
Here we have to note that, as explained in Appendix~\ref{app:surface}, the variation of system size is restricted to homothetic variations. 
For our system, the homothetic variation is the spherical variation due to the spherical symmetry.

\subsection{Choice of $\alpha_w$ of working hypothesis~2}
\label{sec:dst.IE}

This subsection determines the form of $\alpha_w$ which appears in Eq.\eref{eq:action.IE}. 
As required in the working hypothesis~2, $\alpha_w$ is a dimensionless factor composed of the quantity determined at the heat wall. 
This implies that $\alpha_w$ is a function of the parameter,
\eqb
\label{eq:dst.x}
 x \defeq H\,r_w \, ,
\eqe
which means $f_w = 1 - x^2$. 
This $x$ is regarded as the dimensionless quantity determined at the heat wall.
Then our Euclidean action $I_E$ is expressed as
\eqb
 I_E = \frac{\pi}{H^2}\,\left(\, 1 - 2\,x\,f_w - 2\,x\,\alpha_w(x)\,\sqrt{f_w} \,\right) \, .
\eqe
In order to determine $\alpha_w(x)$ so as to preserve the entropy-area law which is the equation of state verified already in the micro-canonical ensemble~\cite{ref:dst.micro,ref:micro}, we need the free energy. 
The free energy $\Fds$ of CEH is obtained via Eq.\eref{eq:app.action.F} of Appendix~\ref{app:action},
\eqb
\label{eq:dst.F_def}
 \Fds(\Tds,\Ads) \defeq 
 - \frac{1}{2 H \sqrt{f_w}}\,\left(\, 1 - 2\,x\,f_w - 2\,x\,\alpha_w(x)\,\sqrt{f_w} \,\right) \, ,
\eqe
where $H$ and $x$ are regarded as functions of $\Tds$ and $\Ads$. 
Then, by the assumption~3, the entropy $\Sds$ of CEH is defined by,
\eqb
\label{eq:dst.S_def}
 \Sds \defeq - \pd{\Fds(\Tds,\Ads)}{\Tds} = - \frac{\partial_H \Fds}{\partial_H \Tds}
 = \frac{\pi}{H^2}\,D\left[\,\alpha_w(x)\,\right] \, ,
\eqe
where
\eqb
 D\left[\,\alpha_w(x)\,\right] \defeq
 - 2\,x^2\,f_w^{3/2}\,\frac{{\rm d}\alpha_w}{{\rm d}x}
 - \left(\, 1 - 2\,x^3 \,\right)\,f_w + x^2 \, .
\eqe
Therefore, to preserve the entropy-area law, $\alpha_w(x)$ has to be a solution of the first-order differential equation, $D\left[\,\alpha_w(x)\,\right] = 1$. 
Then we obtain
\eqb
\label{eq:dst.alphaw}
 \alpha_w(x) = \left(\, \frac{1}{x} - 1 \,\right)\,\sqrt{f_w} + k_w \, ,
\eqe
where $k_w$ is the integration constant. 
The value of $k_w$ can not be determined at present. 
However, as will be shown in Sec.\ref{sec:dst.sigma}, by the existence of surface pressure at a limit $r_w \to 0$, we find $k_w$ is zero,
\eqb
\label{eq:dst.kw}
 k_w = 0 \, .
\eqe
Although the verification of this value is shown later, we proceed our calculations with setting $k_w = 0$ for simplicity of discussion. 
Then, adopting Eq.\eref{eq:dst.kw}, the Euclidean action~\eref{eq:action.IE} is determined,
\eqb
\label{eq:dst.IE}
 I_E = - \dfrac{\pi}{H^2}\,\left( 1 - 2\, x^2 \right) \, .
\eqe
A comment on the limit $I_E \to - \pi/H^2$ as $r_w \to 0$ will be given in Sec.\ref{sec:sd}.

\subsection{Free energy, entropy and second law}
\label{sec:dst.F.S}

Previous subsection has given us the free energy $\Fds$ and the entropy $\Sds$.
The free energy is
\eqb
\label{eq:dst.F}
 \Fds(\Tds,\Ads) = \frac{1 - 2\,x^2}{2\,H\,\sqrt{f_w}} \, ,
\eqe
where $x$ is defined in Eq.\eref{eq:dst.x}. 
As noted at the working hypothesis~1, $\Fds$ should be regarded as a function of $\Tds$ and $\Ads$. 
And the entropy is
\eqb
\label{eq:dst.S}
 \Sds = \frac{\pi}{H^2} \, .
\eqe

Concerning the entropy, it should be noted that, exactly speaking, the second law is not a ``theorem'' proven by some other assumptions, but the basic assumption which can not be proven in the framework of thermodynamics. 
The best way to believe the second law is to ``check'' the statement of the law for as many processes as we can. 
For the black hole thermodynamics, the generalized second law is checked for some representative processes~\cite{ref:gsl}. 
For de~Sitter thermodynamics, we need to check the generalized second law for as many processes as we can. 
However let us expect that the generalized second law holds also for the CEH in de~Sitter spacetime, since our basic assumptions~1,~2, and~3 are the natural extension of consistent black hole thermodynamics.

\subsection{Internal energy}
\label{sec:dst.E}

The internal energy $\Eds$ of CEH is defined by the argument of ordinary statistical mechanics,
\eqb
 \Eds \defeq
  - \left.\pd{\ln Z_{cl}}{(1/\Tds)}\right|_{\Ads=\mbox{const.}}
  = \pd{(\Fds/\Tds)}{(1/\Tds)}
  = \Fds + \Tds \, \Sds
\label{eq:dst.E.2}
\eqe
where Eq.\eref{eq:app.action.F} of Appendix~\ref{app:action} is used at the second equality and the definition of $\Sds$ in Eq.\eref{eq:dst.S_def} is used at the third equality. 
The third equality, $\Eds = \Fds + \Tds \, \Sds$, is the Legendre transformation between $\Fds(\Tds,\Ads)$ and $\Eds$. 
This implies that $\Eds$ is a function of $\Sds$ and $\Ads$, which is consistent with the ordinary thermodynamic argument that the internal energy is a function of extensive state variables. 
Then we get
\eqb
\label{eq:dst.E}
 \Eds(\Sds,\Ads) \defeq \frac{1}{H}\,\sqrt{f_w} \, ,
\eqe
where $H$ and $x$ are regarded as functions of $\Sds$ and $\Ads$ via Eqs.\eref{eq:dst.A},~\eref{eq:dst.x} and~\eref{eq:dst.S}. 
The origin of internal energy $\Eds$ will be discussed in Sec.\ref{sec:sd}.

\subsection{Surface pressure}
\label{sec:dst.sigma}

As explained in Appendix~\ref{app:surface}, the conjugate state variable to $\Ads$ is the surface pressure $\sigmads$ defined by
\eqb
\label{eq:dst.sigma_def}
 \sigmads \defeq - \pd{\Fds(\Tds,\Ads)}{\Ads} =
  - \frac{(\partial_H \Fds)\,(\partial_{r_w} \Tds) - (\partial_{r_w} \Fds)\,(\partial_H \Tds)}
         {(\partial_H \Ads)\,(\partial_{r_w} \Tds) - (\partial_{r_w} \Ads)\,(\partial_H \Tds)} \, ,
\eqe
where Eq.\eref{eq:app.formulas} of Appendix~\ref{app:formulas} is used at the second equality.

As mentioned in Sec.\ref{sec:dst.IE}, we determine the integration constant $k_w$ in this subsection. 
To do so, we calculate $\sigmads$ without setting $k_w$ zero. 
From Eqs.\eref{eq:dst.F_def} and~\eref{eq:dst.alphaw}, the free energy $\tilde{F}$ with non-zero $k_w$ is $\tilde{F} = \Fds + k_w\,r_w$, where $\Fds$ is shown in Eq.\eref{eq:dst.F}. 
Then we obtain from Eq.\eref{eq:dst.sigma_def},
\eqb
 \sigmads = \frac{H}{8\,\pi\,\sqrt{f_w}} - \frac{k_w}{8\,\pi\,r_w} \, .
\eqe
It is obvious that, if $k_w \neq 0$, then the surface pressure diverges in the limit $r_w \to 0$. 
Hence, when we require the existence of finite $\sigmads$ for thermal equilibrium states of CEH with arbitrarily small heat wall, it is natural to set $k_w = 0$. 
Then, with adopting this choice $k_w = 0$, we obtain
\eqb
\label{eq:dst.sigma}
 \sigmads = \frac{H}{8\,\pi\,\sqrt{f_w}} = \frac{1}{4}\,\Tds \, .
\eqe

\subsection{First and third laws}
\label{sec:dst.1st_3rd}

By definitions of $\Sds$ and $\sigmads$, 
\eqb
 d\Fds(\Tds,\Ads) = - \Sds \,d\Tds - \sigmads\,d\Ads \, .
\eqe
Then the first law holds automatically via the Legendre transformation in Eq.\eref{eq:dst.E.2},
\eqb
 d\Eds(\Sds,\Ads) = d(\Fds + \Tds \,\Sds) = \Tds \,d\Sds - \sigmads\,d\Ads \, .
\eqe

Next, to discuss the third law, note that $\Tds$ is monotonically increasing as a function of $r_w$ as shown by $\partial_{r_w} \Tds(H,r_w) = H^3\,r_w/(2\,\pi\,f_w^{3/2}) > 0$. 
The minimum value of $\Tds$ as a function of $r_w$ is given at $r_w = 0$, $\left.\Tds\right|_{r_w\to 0} = H/2\,\pi$. 
This denotes the zero-temperature state is achieved only by the process $H \to 0$ and $r_w \to 0$. 
However it is obvious from Eq.\eref{eq:dst.E} that the infinite energy is required to realize the process $H \to 0$. 
The infinite energy supply is unphysical. 
Hence the third law holds in the sense that the zero-temperature state can not be achieved by any physical process.

\subsection{Scaling law, Euler relation and $\Lambda$ as a ``hidden'' variable}
\label{sec:dst.scaling}

This subsection demonstrates the consistency of the scaling law, which is introduced in the assumption~2, with the other assumptions~1 and~3.

Let us consider the scaling of length size,
\eqb
 r_w \to \lambda\,r_w \quad,\quad
 r_c \to \lambda\,r_c \, .
\eqe
The scaling of CEH radius denotes $H \to \lambda^{-1}\,H$. 
Then from Eqs.\eref{eq:dst.T} and~\eref{eq:dst.sigma}, we get the scaling of intensive variables,
\eqb
 \Tds \to \frac{1}{\lambda}\,\Tds \quad,\quad \sigmads \to \frac{1}{\lambda}\,\sigmads \, .
\eqe
From Eqs.\eref{eq:dst.A} and~\eref{eq:dst.S}, the scaling of extensive variables is
\eqb
 \Ads \to \lambda^2\,\Ads \quad,\quad \Sds \to \lambda^2\,\Sds \, .
\eqe
From Eqs.\eref{eq:dst.F} and~\eref{eq:dst.E}, we get the scaling of thermodynamic functions,
\eqb
 \Fds \to \lambda\,\Fds \quad,\quad \Eds \to \lambda\,\Eds \, .
\eqe
Therefore we find that the scaling law of assumption~2 is consistent with the assumptions~1 and~3. 
Concerning this consistency, the definition of temperature in Eq.\eref{eq:app.action.T} should be emphasized. 
It is obvious that the temperature has the dimension of the inverse of length size by definition. 
Hence, the scaling law of assumption~2 is necessary to adopt Eq.\eref{eq:app.action.T} as the temperature which should be intensive.

Furthermore, to show a more robust consistency of the scaling law, recall that the internal energy is a function of $\Sds$ and $\Ads$. 
Then the above scaling law implies
\eqb
\label{eq:dst.euler.1}
 \lambda\,\Eds(\Sds,\Ads) = \Eds(\lambda^2 \Sds,\lambda^2 \Ads) \, .
\eqe
This denotes that $\Eds(\Sds,\Ads)$ is the homogeneous expression of degree $1/2$. 
By the partial differential of this equation with respect to $\lambda$, we get the Euler relation,
\eqb
 \frac{1}{2}\,\Eds(\Sds,\Ads) = \Tds\,\Sds - \sigmads\,\Ads \, .
\label{eq:dst.euler.2}
\eqe
Note that the concrete functional forms of state variables is not used in deriving this Euler relation.
The Euler relation~\eref{eq:dst.euler.2} is obtained from the scaling behavior~\eref{eq:dst.euler.1} and the differential relations implied by the first law, $\Tds \equiv \partial \Eds(\Sds,\Ads)/\partial \Sds$ and $\sigmads \equiv -\partial \Eds(\Sds,\Ads)/\partial \Ads$. 
On the other hand, we can check that the concrete forms of state variables $\Tds$, $\Sds$, $\sigmads$ and $\Ads$ obtained in previous subsections satisfy the relations~\eref{eq:dst.euler.1} and~\eref{eq:dst.euler.2}. 
Hence the state variables obtained so far are completely consistent with the scaling law of assumption~2.

Finally in this subsection, recall that the cosmological constant $\Lambda$ is regarded as an independent variable in the working hypothesis~1. 
It is obvious that the assumption~2, via the relation $\Lambda = 3\,H^2$, excludes the ``bare $\Lambda$'' from state variables, since $\Lambda$ is neither intensive nor extensive. 
Hence the bare $\Lambda$ can not be a ``state variable'' but a ``hidden variable'' in the consistent de~Sitter thermodynamics.

\subsection{Heat capacity and thermal stability}
\label{sec:dst.C}

The thermodynamic consistency of our de~Sitter canonical ensemble has been clearly checked so far. 
This subsection researches the thermal stability of CEH. 
The appropriate quantity to consider thermal stability is the heat capacity.

The representative heat capacity may be the heat capacity $C_{\Ads}$ at constant $\Ads$. 
Since $\Ads$ depends only on $r_w$, $C_{\Ads}$ describes the response of temperature to the energy supply into CEH with fixing the position of observer $r_w$. 
$C_{\Ads}$ is defined by
\eqb
\label{eq:dst.CA.def}
 C_{\Ads} \defeq \Tds\,\pd{\Sds(\Tds,\Ads)}{\Tds}
  = \Tds\,\frac{\partial_H \Sds}{\partial_H \Tds} \, ,
\eqe
where $\Tds$ and $\Ads$ are regarded as independent variables. 
By Eqs.\eref{eq:dst.T} and~\eref{eq:dst.S}, we get
\eqb
 C_{\Ads} = - \frac{2\,\pi}{H^2}\,f_w \, .
\eqe
Obviously the heat capacity $C_{\Ads}$ is negative definite. 
When the energy is supplied to (extracted from) the CEH, then the temperature $\Tds$ decreases (increases). 
This denotes the CEH is thermally unstable. 
However, when we consider the constant $\Lambda$ process as the physical one, this heat capacity $C_{\Ads}$ is the capacity for unphysical process, since $C_{\Ads}$ is defined by the derivative with respect to $H$ as seen in Eq.\eref{eq:dst.CA.def}. 
The thermal instability due to negative $C_{\Ads}$ seems to be unphysical.

When we consider the constant $\Lambda$ process as the physical one, the heat capacity $C_{\Lambda}$ at constant $\Lambda$ is of interest. 
It is defined by
\eqb
 C_{\Lambda} \defeq \Tds\,\pd{\Sds(\Tds,H)}{\Tds}
  = \Tds\,\frac{\partial_{r_w} \Sds}{\partial_{r_w} \Tds} \, ,
\eqe
where $\Tds$ and $\Lambda$ are regarded as independent variables. 
Then, since $\Sds$ is independent of $r_w$, we find
\eqb
\label{eq:dst.CLambda}
 C_{\Lambda} = 0 \, .
\eqe
Since $C_{\Lambda}$ is not negative but \emph{zero}, thermal equilibrium of CEH is not thermally unstable but thermally \emph{marginal} stable for constant $\Lambda$ process.

Since the constant $\Lambda$ process means the variation of only the position of observer $r_w$ with fixing the CEH radius $r_c = \sqrt{3/\Lambda}$, the vanishing heat capacity~\eref{eq:dst.CLambda} means that the observer's position $r_w$ can change without a heat supply to the CEH. 
Here note that, Eq.\eref{eq:dst.CLambda} does not imply that changing $r_w$ has no thermodynamic effect on the CEH. 
For example, changing $r_w$ in constant $\Lambda$ process gives rise to the change of surface pressure, $\partial \sigmads/\partial r_w \neq 0$. 
The vanishing heat capacity at constant $\Lambda$ process~\eref{eq:dst.CLambda} means simply the disappearance of heat supply in changing $r_w$ with fixing $\Lambda$. 
This is a peculiar thermodynamic property of the CEH.

\subsection{Surface compressibility and mechanical stability}
\label{sec:dst.kappa}

Let us research the mechanical stability of our thermal equilibrium system of CEH. 
As explained in Appendix~\ref{app:surface}, the appropriate quantity to consider the mechanical stability may be the isothermal surface compressibility $\kappa_{\Tds}$ defined by $\kappa_{\Tds} \defeq \Ads^{-1}\,\partial \Ads(\Tds,\sigmads)/\partial \sigmads$. 
However, since $\sigmads$ is proportional to $\Tds$ as shown in Eq.\eref{eq:dst.sigma}, the definition of $\kappa_{\Tds}$ becomes meaningless. 
Then, instead of $\kappa_{\Tds}$, let us consider the ``isentropic'' surface compressibility $\kappa_{\Sds}$ defined by
\eqb
 \kappa_{\Sds} \defeq \frac{1}{\Ads}\,\pd{\Ads(\Sds,\sigmads)}{\sigmads} \, ,
\eqe
where $\Sds$ and $\sigmads$ are regarded as independent variables.
Since $\Sds$ depends only on $\Lambda$ as shown in Eq.\eref{eq:dst.S}, $\kappa_{\Sds}$ is equivalent to the surface compressibility at constant $\Lambda$, and therefore it seems to be the physical quantity. 
By Eqs.\eref{eq:dst.A},~\eref{eq:dst.S} and~\eref{eq:dst.sigma}, we get
\eqb
 \kappa_{\Sds} = \frac{1}{\Ads}\,\frac{\partial_{r_w}\Ads}{\partial_{r_w}\sigmads}
  = \frac{16\,\pi\,f_w^{3/2}}{H\,x^2} \, .
\eqe
Obviously $\kappa_{\Sds}$ is positive definite. 
When the surface $\Ads$ increases, the surface pressure $\sigmads$ also increases. 
If we take the same criterion of mechanical stability as York~\cite{ref:euclidean.sch} (see the end of Appendix~\ref{app:surface}), then the positivity of $\kappa_{\Sds}$ implies that our thermal equilibrium system is mechanically stable.

\section{Summary and discussions}
\label{sec:sd}

Referring to the ``thermodynamically consistent'' black hole canonical ensemble formulated by York~\cite{ref:euclidean.sch}, we have introduced the basic assumptions~1,~2 and~3 to construct the thermodynamically consistent de~Sitter canonical ensemble. 
The need of those assumptions was discussed in Sec.\ref{sec:ass.preliminary}. 
The central technique is the Euclidean action method introduced by Gibbons and Hawking~\cite{ref:euclidean}, in which the choice of integration constant is important. 
The integration constant in our Euclidean action~\eref{eq:dst.IE} has been determined so as to preserve the entropy-area law which is the equation of state verified already in the micro-canonical ensemble~\cite{ref:dst.micro,ref:micro}. 
Then as shown in Sec.\ref{sec:dst}, we have constructed the thermodynamically consistent de~Sitter canonical ensemble, which describes a stable equilibrium state as discussed in Secs.\ref{sec:dst.C} and~\ref{sec:dst.kappa} with the help of Appendix~\ref{app:surface}. 
It is also found in Sec.\ref{sec:dst.scaling} that, due to the assumption~2, when $\Lambda$ is regarded as a working variable, the bare $\Lambda$ can not be a state variable but a ``hidden variable'' in the consistent de~Sitter thermodynamics.

As mentioned in the working hypothesis~1, we regard $\Lambda$ as a working variable to obtain thermodynamically consistent de~Sitter canonical ensemble. 
The validity of this working hypothesis~1 can be recognized simply by the following fact: 
The entropy $\Sds = 3 \pi/\Lambda$ depends only on $\Lambda$ as already verified in the micro-canonical ensemble~\cite{ref:dst.micro,ref:micro}, and consequently the definition of $\Sds$ in Eq.\eref{eq:dst.S_def} is expressed by using the derivatives of $\Fds$ and $\Tds$ with respect to $\Lambda$. 
The derivative with respect to $\Lambda$ requires implicitly the variable $\Lambda$. 
Hence, in order to calculate the entropy in the canonical ensemble, it is necessary to adopt the working hypothesis~1. 
\emph{The canonical ensemble of de~Sitter spacetime constructs the ``generalized'' thermodynamics in which $\Lambda$ behaves as a working variable, and the physical process is described by the constant $\Lambda$ process.}

Note that, in de~Sitter micro-canonical ensemble proposed by Hawking and Ross~\cite{ref:dst.micro}, the number of states $W$ is equal to $\exp I_E^{\rm (micro)}$, where $I_E^{\rm (micro)}$ is the Euclidean action of de~Sitter space without heat wall. 
Because the heat wall is not introduced and there is no boundary in the Euclidean de~Sitter space for the micro-canonical ensemble, $I_E^{\rm (micro)}$ consists of only the bulk term (first term) of $I_L$ and we obtain $I_E^{\rm (micro)} = \pi\,r_c^2$. 
(Then, as mentioned in Sec.\ref{sec:intro}, the Boltzmann's relation yields the entropy-area law, $\Sds \defeq \ln W = \pi r_c^2$.) 
On the other hand, as shown in this paper, the partition function of de~Sitter canonical ensemble is the Euclidean action $I_E$ obtained in Eq.\eref{eq:dst.IE}. 
When we consider an arbitrarily small heat wall $r_w \to 0$, the action $I_E$ takes the limit value $I_E \to - \pi r_c^2$ as $r_w \to 0$. 
This limiting value is different from $I_E^{\rm (micro)}$ by the negative signature. 
Here one may naively think that our $I_E$ should coincide with $I_E^{\rm (micro)}$ at this limit. 
This naive requirement seems reasonable from the point of view of spacetime geometry, but is not necessarily reasonable from the point of view of statistical mechanics, because the coincidence of $I_E$ with $I_E^{\rm (micro)}$ at the limit $r_w \to 0$ means that the micro-canonical ensemble is some limiting case of the canonical ensemble. 
In statistical mechanics, the micro-canonical ensemble is not some limiting case of the canonical ensemble. 
In de~Sitter thermodynamics, the limit $r_w \to 0$ is just the case of an arbitrarily small heat wall and not the case without heat wall. 
Therefore, in statistical mechanical sense, there seems to be no reason to require that $I_E$ coincides with $I_E^{\rm (micro)}$ at the limit $r_w \to 0$. 
(Concerning this discussion, see also the end of Sec.\ref{sec:ass.sch} in which the limiting case of large heat bath for black hole thermodynamics is summarized.)

Finally let us discuss the origin of internal energy $\Eds$. 
For the first, recall the Schwarzschild thermodynamics formulated by York~\cite{ref:euclidean.sch}. 
The internal energy $E_{\rm BH}$ in Schwarzschild thermodynamics is related to its mass parameter $M$ by
\eqb
 M = E_{\rm BH} - \frac{E_{\rm BH}^{\quad 2}}{2\,r_w} \,,
\eqe
where $r_w$ is the outer-most radius of cavity shown in Fig.\ref{fig:1}. 
The second term $E_{\rm BH}^{\quad 2}/(2 r_w)$ can be interpreted as the self-gravitational potential energy of black hole. 
Then $E_{\rm BH}$ is interpreted as the ``bare'' mass energy of the \emph{black hole in cavity}, while $M$ is the ``net'' mass energy including the self-gravitational potential. 
It seems reasonable to consider that the origin of internal energy $E_{\rm BH}$ is the mass of black hole. 
The mass $M$ as the origin of energy $E_{\rm BH}$ can be clearly exhibited in the large cavity limit $r_w \to \infty$. 
In this limit we have $E_{\rm BH}\,|_{r_w\to\infty} = M$, which manifestly shows that $E_{\rm BH}$ is originated from $M$.

Then turn our discussion to de~Sitter thermodynamics. 
Let us consider the small heat wall limit $r_w \to 0$, which seems to correspond to the large cavity limit in Schwarzschild thermodynamics, since the heat wall is most distant from CEH. 
In this limit we have
\eqb
 \lim_{r_w \to 0} \Eds = \frac{1}{H} \,.
\eqe
This may show that the origin of internal energy $\Eds$ is the cosmological constant $\Lambda\,(= 3 H^2 )$. 
In the framework of classical general relativity, the de~Sitter spacetime is a vacuum spacetime which includes no energy source. 
However, in the de~Sitter thermodynamics which includes essentially the quantum gravitational effects, $\Lambda$ may be interpreted as a kind of energy source which is responsible to the energy $\Eds$. 
Also $\Lambda$ may be responsible to the entropy $\Sds$.

\section*{Acknowledgments}

I'd like to thank my colleague Toshihide Futamura, a mathematician, for his help to notice useful formulas displayed in Appendix~\ref{app:formulas}. 
This work is supported by the Grant-in-Aid for Scientific Research Fund of the Ministry of Education, Culture, Sports, Science and Technology, Japan (Young Scientists (B) 19740149).

\appendix
\section{Euclidean action as a partition function of spacetime}
\label{app:action}

The Euclidean action method for systems including gravity is originally introduced by Gibbons and Hawking~\cite{ref:euclidean} in an analogy with the thermal field theory of matter fields in flat spacetime. 
For the first, thermal fields in flat spacetime is summarized. 
Then its generalization by Gibbons and Hawking is briefly reviewed.

\subsection{Thermal fields in flat spacetime}

The thermal field theory is the statistical mechanics of quantum fields in thermal equilibrium~\cite{ref:tft}. 
The partition function for the canonical ensemble of a field $\phi$ in Minkowski spacetime is defined by the path integral,
\eqb
 Z_{\rm flat} \defeq \int {\mathcal D}\phi\,e^{I_E[\phi]} \, ,
\label{eq:app.action.Zflat}
\eqe
where ${\mathcal D}\phi$ is a normalized measure of path integral and $I_E[\phi]$ is the Euclidean action of $\phi$ defined by
\eqb
 I_E[\phi] \defeq i\times \mbox{Lorentzian action with replacing $t$ by $-i\,\tau$} \, ,
\label{eq:app.action-IE.flat}
\eqe
where the Lorentzian metric signature is $(- + + +)$, the time coordinate $t$ in the Minkowski spacetime is of ordinary Cartesian coordinates (the time-time component of metric is $-1$), and the replacement of real time $t$ by imaginary time $\tau$ is called the \emph{Wick rotation}. 
By the Wick rotation $t \to - i\, \tau$, the metric in evaluating $I_E[\phi]$ becomes that of flat Euclidean space with signature $(+ + + +)$. 
The ``direction'' of Wick rotation on complexified time plane is ``clockwise'' $t \to -i\,\tau$ (not ``counterclockwise'' $t \to + i\,\tau$) in order to make $Z_{\rm flat}$ correspond to the partition function of (grand-)canonical ensemble in quantum statistics~\cite{ref:tft}. 
In the path integral in Eq.\eref{eq:app.action.Zflat}, an appropriate boundary condition is also given to $\phi$ in order to realize a thermal equilibrium state. 
At least, because thermal equilibrium state is static, a periodic boundary condition in the imaginary time direction is required,
\eqb
 \phi(\tau) = \phi(\tau + \beta) \, ,
\eqe
where $\beta$ is the imaginary time period~\footnote{
When the periodic boundary condition in imaginary time is not required, the path integral in Eq.\eref{eq:app.action.Zflat} describes an ordinary transition amplitude of $\phi$ in ordinary quantum field theory of zero temperature.
}. 
With this condition, it has already been known~\cite{ref:tft} that $Z_{\rm flat}$ corresponds to the partition function of canonical ensemble of equilibrium temperature $T_{\rm flat}$ defined by
\eqb
 T_{\rm flat} \defeq \frac{1}{\beta} \, .
\eqe
$Z_{\rm flat}$ describes thermal equilibrium state of $\phi$ of equilibrium temperature $T_{\rm flat}$ in Minkowski spacetime, and the free energy $F_{\rm flat}$ of the equilibrium state is obtained,
\eqb
 F_{\rm flat} = -T_{\rm flat}\,\ln Z_{\rm flat} \, .
\eqe

\subsection{Curved spacetime and thermal fields on it}

In curved spacetime, we consider a thermal equilibrium state of the combined system of spacetime and matter field. 
For the canonical ensemble of our combined system, it is usually assumed that the partition function $Z$ is obtained by replacing flat metric in Eq.\eref{eq:app.action.Zflat} with curved one~\cite{ref:euclidean},
\eqb
 Z \defeq \int {\mathcal D}g_E \cdot {\mathcal D}\phi\,e^{I_E[g_E,\phi]} \, ,
\label{eq:app.action-Z}
\eqe
where
\eqb
 I_E[g_E,\phi] \defeq
 i\times \mbox{$I_L[g,\phi]$ with Wick rotation $t \to -i\,\tau$} \, ,
\label{eq:app.action.IE.curved}
\eqe
where $g_E$ is the Euclidean metric of signature $(+ + + +)$ obtained from the Lorentzian metric $g$ by the Wick rotation $t \to - i \tau$, and $I_L[g,\phi]$ is the sum of Lorentzian Einstein-Hilbert action and Lorentzian matter action. 
Here since the spacetime metric $g$ is also assumed to be quantum metric, $g_E$ appears as an integral variable in the path integral~\eref{eq:app.action-Z}. 
To consider equilibrium states of spacetime with matter field, the periodic boundary condition in imaginary time is required for not only $\phi$ but also $g_E$,
\eqb
g_{E\,\mu \nu}(\tau) = g_{E\,\mu \nu}(\tau + \beta) \, .
\eqe
Here the equilibrium temperature of $g$ and $\phi$ is not defined simply by $\beta^{-1}$, because the spacetime is curved. 
Instead of the simple inverse $\beta^{-1}$, the temperature $T$ should be defined by the proper length in the Euclidean space of $g_E$ in the imaginary time direction,
\eqb
 T \defeq \left[\, \int_0^{\beta}\,\sqrt{g_{E\,\tau \tau}}\,d\tau \,\right]^{-1} \, .
\label{eq:app.action.T}
\eqe
Since the metric component $g_{E\,\tau \tau}$ is a function of spacetime coordinates, the integral in Eq.\eref{eq:app.action.T} becomes a function of spatial coordinates. 
Therefore it is important to specify where the temperature is defined. 
Here note that the Euclidean action method is for the canonical ensemble. 
This implies the existence of a heat bath whose temperature coincides with the temperature of the system under consideration, since the system is in a thermal equilibrium with the heat bath. 
Therefore it is reasonable to evaluate $g_{E\,\tau \tau}$ in Eq.\eref{eq:app.action.T} at the contact surface of the system with the heat bath. 
The contact surface is the boundary of the spacetime region. 
Hence the temperature $T$ should be evaluated at the spacetime boundary.

In the path integral in Eq.\eref{eq:app.action-Z}, the metric and matter field are not necessarily solutions of classical Einstein equation and field equations. 
However when the field $\phi$ is weak enough, the dominant contribution would come from the classical solutions, $g_{cl}$ and $\phi_{cl}$, and we can expand as
\eqb
\label{eq:app.expand.g}
 g_{\mu\nu} = g_{cl\,\mu\nu} + \delta g_{\mu\nu} \quad,\quad
 \phi = \phi_{cl} + \delta \phi \, ,
\eqe
where $\delta g$ and $\delta \phi$ describe quantum/statistical fluctuations of metric and matter. 
In spacetimes with event horizon, this expansion seems reasonable since the Hawking temperature is usually very low and the matter field $\phi$ of Hawking radiation is weak. 
Then the Euclidean action becomes
\eqb
 I_E[g_E,\phi] = I_E[g_{E\,cl},\phi_{cl}] + I_E[\delta g_E] + I_E[\delta \phi]
             + \mbox{higher order terms} \, ,
\label{eq:app.expand.IE}
\eqe
where $g_{E\,cl}$ is the Euclidean metric obtained from $g_{cl}$, and the second and third terms are quadratic in fluctuations by definition of classical field equations. 
The partition function becomes
\eqb
 \ln Z = I_E[g_{E\,cl},\phi_{cl}] + \ln\int \mathcal{D}(\delta g_E)\,e^{I_E[\delta g_E]}
         + \ln\int \mathcal{D}(\delta \phi)\,e^{I_E[\delta \phi]} + \cdots \, .
\eqe
The leading term $I_E[g_{E\,cl},\phi_{cl}]$ includes only the classical solutions. 
Hence the partition function $Z_{cl}$ of the thermal equilibrium state of background classical spacetime and matter is defined by
\eqb
 \ln Z_{cl} \defeq I_E[g_{E\,cl},\phi_{cl}] \, .
\label{eq:app.action.Zcl}
\eqe
The state variables obtained from $Z_{cl}$ describe thermal equilibrium states of background spacetime and matter. 
For spacetimes with event horizon, $Z_{cl}$ is interpreted as the partition function of the event horizon. 
For empty background spacetimes ($\phi_{cl} = 0$) like Schwarzschild and de~Sitter spacetimes, $Z_{cl}$ is determined by only classical metric, $\ln Z_{cl} = I_E[g_{E\,cl}]$. 
This $I_E[g_{E\,cl}]$ describes the canonical ensemble of the thermal equilibrium states of the background classical spacetime, where the thermal equilibrium is achieved by the interaction with the quantum fluctuations of metric. 
Then the free energy of those classical background spacetimes are determined by
\eqb
 F = -T\,\ln Z_{cl} = - T\,I_E[g_{E\,cl}] \, ,
\label{eq:app.action.F}
\eqe
where $T$ is defined by Eq.\eref{eq:app.action.T} with replacing $g_E$ by $g_{E\,cl}$. 
This $T$ is the equilibrium temperature of the event horizon.

\section{Surface area, surface pressure and mechanical stability}
\label{app:surface}

In the consistent thermodynamic formulation of Schwarzschild black hole by York~\cite{ref:euclidean.sch}, the surface area of heat bath is the consistent state variable of system size (see Fig.\ref{fig:1}). 
As shown in Sec.\ref{sec:ass}, the consistency of the surface area of heat bath as a state variable is also retained in our de~Sitter thermodynamics. 
However for the ordinary laboratory systems, it is not the area but the volume which behaves as the consistent state variable of system size. 
Therefore, for the purpose of a consistent formulation of de~Sitter thermodynamics (and also of black hole thermodynamics), it is necessary to summarize the physical meanings of the surface area and the thermodynamic conjugate state variable to it.

The physical meaning of surface area (and surface pressure and isothermal surface compressibility defined below) may not be clearly explained in the original work of consistent black hole thermodynamics by York~\cite{ref:euclidean.sch}. 
We expect this appendix may be helpful to consider not only de~Sitter thermodynamics but also black hole thermodynamics.

In this appendix we consider an ordinary laboratory system of volume $V_o$ and surface area $A_o$. 
If the system is a gas in a container, $V_o$ and $A_o$ are respectively the volume and surface area of the container. 
The state variable of system size is $V_o$, which is extensive by definition. 
In general, the set of numerical values of independent state variables are uniquely determined to each thermal equilibrium state. 
The independent state variables are regarded as the ``coordinate'' in the state space of thermal equilibrium states. 
For example, the numerical value of the set $(T_o,V_o)$ at a thermal equilibrium state is different from that at the other thermal equilibrium state, where $T_o$ is the temperature of the system under consideration. 
However, the surface area $A_o$ can not be the state variable, because we can deform the geometrical shape of the system with keeping the ``coordinate'' $(T_o,V_o)$ unchanged. 
Furthermore, while $V_o$ is the extensive variable, $A_o$ is never extensive since $A \to \nu^{2/3} A_o$ under the volume scaling $V_o \to \nu V_o$.

However if the variation of the system size is restricted to the \emph{homothetic variation}, there is one-to-one correspondence between $V_o$ and $A_o$. 
For example when the system is spherically symmetric, the one-to-one correspondence is realized under the spherical variation of that system. 
Hence, if the variation of system size is homothetic, the surface area $A_o$ instead of the volume $V_o$ is regarded as a kind of state variable of the system size which has irregular scaling behavior for ordinary laboratory systems. 
However for the black hole and de~Sitter thermodynamics as explained in Sec.\ref{sec:ass}, the extensive variables $X$ are scaled as $X \to \nu^{2/3}\, X$ under the volume scaling of the scaling rate $\nu$. 
Therefore the surface area $A_o$ for black hole and de~Sitter CEH seems to be the extensive state variable of system size which possesses consistent scaling property (see Sec.\ref{sec:dst.scaling}).

Next, let us discuss about the state variable which is thermodynamically conjugate to the system size. 
When the volume $V_o$ is chosen as the state variable of the size of ordinary laboratory systems, the conjugate intensive state variable to $V_o$ is the pressure $P_o$. 
It can be defined via the free energy $F_o(T_o,V_o)$ of the system, $P_o \defeq - \partial F_o(T_o,V_o)/\partial V_o$. 
Keeping this relation in mind, for the case that the surface area $A_o$ is chosen as the state variable of system size, let us define a state variable $\sigma_o$ conjugate to $A_o$ as
\eqb
 \sigma_o \defeq - \left.\pd{F_o(T_o,A_o)}{A_o}\right|_{Homo} \, ,
\label{eq:app.surface.sigma}
\eqe
where $F_o(T_o,A_o) \defeq F_o(T_o,V_o(A_o))$ with regarding $V_o$ as a function of $A_o$, and the restriction ``$Homo$'' means that the variation of system size is restricted to the homothetic ones.
Let us call $\sigma_o$ the \emph{surface pressure} of the system after York~\cite{ref:euclidean.sch}.
Then we get from the definition of the surface pressure,
\eqb
 \sigma_o = - \left.\pd{F(T_o,V_o)}{V_o}\,\frac{d V_o(A_o)}{dA_o}\right|_{Homo}
          = P_o\,L_o \, ,
\eqe
where $L_o \defeq \left.dV_o/dA_o\right|_{Homo}$. 
Here $L_o$ is the length scale determined by the volume variation per unit area variation in the homothetic way, and we call $L_o$ the \emph{homothetic scaling rate}. 
Therefore, the surface pressure $\sigma_o$ is the work done by the system onto its environment per unit surface area.

Here note that $\sigma_o$ is not the surface ``tension'' of a membrane. 
To explain it, recall that there are two versions of the ordinary surface tension. 
One is for the membrane system whose system size is spatially two dimensional, and another is the interface between two phases in a phase equilibrium system. 
The former example is a rubber sheet, and the latter example is the surface of a drop of water in a phase equilibrium with the surrounding vapor. 
Let us discuss the difference of $\sigma_o$ from the surface tensions of these two examples: 
For the first, in thermodynamics of rubber sheet, the system size is definitely the area of the sheet and its tension is defined by the differential of its free energy by the area. 
However our system in this appendix is not the membrane like a rubber sheet but the ordinary gas in a container of ``volume'' $V_o$. 
Hence, although $\sigma_o$ has the same dimension as the surface tension, the thermodynamic meaning of $\sigma_o$ is not the same with the surface tension of membrane systems.

Next, in thermodynamics of the water drop in a phase equilibrium with surrounding vapor, the surface tension $\sigma_{\rm int}$ of the interface between water drop and surrounding vapor is related to the difference between pressures of water and vapor via the \emph{Laplace relation}, $P_{\rm water} - P_{\rm vapor} = 2\,\sigma_{\rm int}/L_{\rm int}$, where $L_{\rm int}$ is the radius of drop (with assuming the spherical symmetry of drop for simplicity). 
If $P_{\rm vapor} = 0$, then one may think $\sigma_o$ is similar to $P_{\rm water}$ under the identification of $L_{\rm int}$ with $2 L_o$. 
However it should be emphasized that the \emph{phase equilibrium} is necessary to obtain the Laplace relation (see for example Kondepudi and Prigogine~\cite{ref:sm}). 
Our system, the gas in a container, is the system of one phase and not in phase equilibrium. 
Furthermore, the gas of our system is in a mechanical equilibrium with the material composing the container which has a \emph{non-zero} pressure. 
Therefore, exactly speaking, the interpretation of $\sigma_o$ as the surface tension, which is based on the equilibrium of two phases, seems to be inappropriate. 
Although the similarity between $\sigma_o$ and the surface tension may be recognized as mentioned above, we do not call $\sigma_o$ the surface tension, but call it the \emph{surface pressure}.

From the above we recognize that the implication of $\sigma_o$ about mechanical stability of the system should be researched in a way independent of the surface tension. 
One may consider a mechanical stability of the system via the surface pressure in an analogy with ordinary pressure. 
When $(T_o , P_o)$ are chosen as independent state variables, the state variable useful to consider the mechanical stability is the isothermal compressibility, $\kappa_{T_o}^{(V_o)} \defeq - P_o^{-1}\,\partial V_o(T_o,P_o)/\partial P_o$, where $V_o$ is regarded as a function of $T_o$ and $P_o$ via the equations of states. 
Then the mechanical stability is described by the positivity, $\kappa_{T_o}^{(V_o)} > 0$. 
In an analogy with $\kappa_{T_o}^{(V_o)}$, we define the \emph{isothermal surface compressibility},
\eqb
 \kappa_{T_o}^{(A_o)} \defeq
 \left.\frac{1}{A_o}\,\pd{A_o(T_o,\sigma_o)}{\sigma_o}\right|_{Homo} \, ,
\label{eq:app.surface.kappa}
\eqe
where $A_o$ is regarded as a function of $T_o$ and $\sigma_o$ via the equations of states. 
Here, following York's black hole thermodynamics~\cite{ref:euclidean.sch}, we take the signature $+$ in the right-hand side. 
Since Eq.\eref{eq:app.surface.kappa} is just a formal definition in an analogy with ordinary one $\kappa_{T_o}^{(V_o)}$, it is not clear at present if the mechanical stability of the system is described by the positivity $\kappa_{T_o}^{(A_o)} > 0$ or not~\footnote{
If the surface pressure is a kind of the ordinary surface tension of membrane systems, then the positivity $\kappa_{T_o}^{(A_o)} > 0$ means the system is mechanically stable.
}.

To get some insight into the mechanical stability, let us examine two examples, the ideal gas and the radiation field in a spherical container of radius $r_o$. 
Due to the spherical symmetry, the homothetic variation is the spherical variation and the homothetic scaling rate becomes, $L_o = r_o/2$. 
For the first consider the ideal gas. 
Its equations of states are $P_o\,V_o = N_o\,T_o$ and $E_o = N_o\,C_v\,T_o$, where $N_o$ is the number of molecules, $E_o$ is the internal energy and $C_v$ is the specific heat at constant volume.
Then we get the ordinary isothermal compressibility $\kappa_{T_o}^{(V_o)} = P_o^{-1} > 0$, which denotes the ideal gas is mechanically stable. 
On the other hand, we obtain the surface pressure $\sigma_o = (3/2)\,N_o\,T_o/A_o > 0$, and the isothermal surface compressibility $\kappa_{T_o}^{(A_o)} = - \sigma_o^{-1} < 0$. 
Therefore, one may think that the mechanical stability of any system is described by $\kappa_{T_o}^{(A_o)} < 0$. 
However it is not always true as shown below.

The second example is the radiation field in a spherical cavity of radius $r_o$. 
Its equations of states are $P_o\,V_o = E_o/3$ and $E_o = a\,T_o^4\,V_o$, where $a$ is the radiation density constant. 
When we consider the ordinary isothermal compressibility $\kappa_{T_o}^{(V_o)}$, the set $(T_o , P_o)$ should be the independent state variables. 
However, because of the relation $P_o = a\,T_o^4/3$ implied by the equations of states, $T_o$ can not be independent of $P_o$, and consequently $\kappa_{T_o}^{(V_o)}$ can not be defined for the radiation field. 
But it is usually expected that the radiation field is one of the thermodynamically normal and mechanically stable systems. 
On the other hand, we can define $\sigma_o$ and $\kappa_{T_o}^{(A_o)}$ for the radiation field. 
They are $\sigma_o = a\,r_o\,T_o^4/6 > 0$ and $\kappa_{T_o}^{(A_o)} = 2/\sigma_o > 0$. 
The isothermal surface compressibility for radiation field has opposite signature to that of ideal gas. 
Hence we can not definitely determine the criterion of mechanical stability by the signature of $\kappa_{T_o}^{(A_o)}$.

However in the black hole thermodynamics formulated by York~\cite{ref:euclidean.sch}, referring to the radiation field, it is assumed that the positivity of isothermal surface compressibility becomes the criterion of mechanical stability of the systems for which the ordinary isothermal compressibility is not defined. 
For the thermal equilibrium of black hole with heat bath (see Fig.\ref{fig:1}) , the ordinary isothermal compressibility can not be defined, because the volume of the system can not be defined uniquely by the same reason as de~Sitter spacetime (see Sec.\ref{sec:dst.A}). 
Then York~\cite{ref:euclidean.sch} suggests that the thermal equilibrium state of black hole with heat bath is mechanically stable (unstable) for sufficiently small (large) heat bath by showing the positivity (negativity) of the isothermal surface compressibility.

\section{Useful differential formulas}
\label{app:formulas}

This appendix shows the differential formula used in Sec.\ref{sec:dst}. 
Let $F$ be a function of $g$ and $h$, $F = F(g,h)$. 
And consider the case that $g$ and $h$ are also functions of $x$ and $y$, $g = g(x,y)$ and $h = h(x,y)$. 
Then define $F(x,y)$ by
\eqb
 F(x,y) \defeq F(\,g(x,y),h(x,y)\,) \, .
\eqe
Let us aim to express the partial derivatives $\partial_g F(g,h)$ and $\partial_h F(g,h)$ by using the derivatives with respect to~$x$ and~$y$. 
By standard differential calculus, we get
\eqb
\begin{split}
 \pd{F(x,y)}{x} &= \pd{F(g,h)}{g}\,\pd{g(x,y)}{x} + \pd{F(g,h)}{h}\,\pd{h(x,y)}{x} \, , \\
 \pd{F(x,y)}{y} &= \pd{F(g,h)}{g}\,\pd{g(x,y)}{y} + \pd{F(g,h)}{h}\,\pd{h(x,y)}{y} \, .
\end{split}
\eqe
Regarding these equations as the algebraic ones of $\partial_g F(g,h)$ and $\partial_h F(g,h)$, we obtain the formulas,
\eqb
\label{eq:app.formulas}
\begin{split}
 \pd{F(g,h)}{g} &=
   \frac{(\partial_x F)\,(\partial_y h) - (\partial_y F)\,(\partial_x h)}
        {(\partial_x g)\,(\partial_y h) - (\partial_y g)\,(\partial_x h)} \, , \\
 \pd{F(g,h)}{h} &=
   \frac{(\partial_x F)\,(\partial_y g) - (\partial_y F)\,(\partial_x g)}
        {(\partial_x h)\,(\partial_y g) - (\partial_y h)\,(\partial_x g)} \, .
\end{split}
\eqe


\end{document}